\documentclass[aps,prb,floatfix,twocolumn]{revtex4}
\usepackage[dvips]{graphicx} 
\usepackage{amsmath}
\usepackage{epsfig}

\begin{document}
\title{Carrier induced ferromagnetism in concentrated and diluted local-moment systems}

\author{ W. Nolting$^{a}$, T. Hickel$^{a}$, A. Ramakanth$^{b}$, G. G. Reddy$^{b}$ and M. Lipowczan$^{c}$}
\affiliation{$^{a}$Institut f{\"u}r Physik, Humboldt-Universit{\"a}t zu
  Berlin, 12489 Berlin, Germany\\$^{b}$Kakatiya University, Department of
  Physics,Warangal-506009, India\\$^{c}$Institute of Physics, Silesian
  University, 40-007 Katowice, Poland}

 
\begin{abstract}
For modeling the magnetic properties of concentrated and diluted
magnetic semiconductors, we use the Kondo-lattice model. The magnetic
phase diagram is derived by inspecting the static susceptibility of itinerant
band electrons, which are exchange coupled to localized magnetic
moments. It turns out that rather low band occupations favour a
ferromagnetic ordering of the local moment systems due to an indirect
coupling mediated by a spin polarization of the itinerant charge carriers. 
The disorder in diluted systems is treated by adding a CPA-type concept
to the theory. For almost all moment concentrations $x$, ferromagnetism is
possible, however, only for carrier concentrations $n$ distinctly smaller than
$x$. The charge carrier compensation in real magnetic semiconductors
(in Ga$_{1-x}$Mn$_{x}$As by e.g. antisites) seems to be a necessary
condition for getting carrier induced ferromagnetism.

\end{abstract}

\maketitle

\section{Introduction}
The exciting research field \textit{``spintronics''} refers to new
phenomena of electronic transport, for which the electron spin plays a
decisive role, in contrast to conventional electronics for which the
electron spin is practically irrelevant. For a full exploitation of
spintronics, one should have materials that are
simultaneously semiconducting and ferromagnetic.That is the reason for
the intensive effort that has been focused on the search for magnetic
semiconductors with high Curie temperatures. It is to the merit of Ohno
and coworkers\cite{OhnoScience,MatsukuraOhno} to reach a $T_{C}$ of up
to $110K$ in Ga$_{1-x}$Mn$_{x}$As and to demonstrate the
electric control of $T_{C}$ by means of a gate voltage
\cite{OhnoNature}. (Even larger $T_{C}$ values have been observed for 
annealed multilayers \cite{CTMO03}.)
Intense experimental as well as theoretical research
on the outstanding phenomena associated with the interplay between
ferromagnetic cooperative features and semiconducting properties is
currently going on \cite{Dietl02}. It is the important challenge of
materials science to understand the ferromagnetism in compounds such as
Ga$_{1-x}$Mn$_{x}$As, and to find out the conditions for Curie temperatures
$T_{C}$  sufficiently  exceeding room temperature. This paper shall
contribute to the fundamentals of ferromagnetism in diluted
local-moment systems.

It is commonly accepted \cite{Dietl02} that the \textit{(ferromagnetic)
Kondo-lattice model (KLM)}, certainly better denoted as  \textit{s-f} or
\textit{s-d model} or, 
in its strong-coupling regime, as \textit{double exchange model},
represents a good starting point for the description of the so-called
local-moment magnetism. To this class of magnetic materials belong the
classical magnetic semiconductors (insulators) such as the Eu
chalcogenides EuO, EuS, EuTe \cite{Wachter79}, which today are classified
as \textit{``concentrated''} magnetic semiconductors. Other
representatives are the local-moment metals Gd, Dy, Tb $\cdots$ as well as
Eu$_{1-x}$Gd$_{x}$S, $\cdots$ , for which magnetic and electrical properties
are provoked by two different electronic subsystems. Strictly localized
$4f$ electrons of the rare earth ion provide the magnetic moment while
itinerant $5d/6s$ electrons take care of the electrical
conductivity. These local-moment systems reveal an exceptionally rich
variety of physical properties with basic ingredients being the electronic
correlations and spin ordering. Thereby, an interband exchange between 
the local moments and the itinerant conduction
electrons appears to play a dominant role, in particular, as far as 
the magnetic and magnetooptic properties are concerned.

The same holds for the already mentioned \textit{diluted magnetic semiconductors
(DMS)}. The implantation of Mn$^{2+}$ ions in the prototypical
semiconductor GaAs provides local moments ($S=\frac{5}{2}$) which
decisively influence the electronic GaAs states giving them, e.g., an
extraordinary temperature dependence. Furthermore, each divalent Mn ion
creates in principle one valence band hole. The temperature
dependence of the band states induced by exchange coupling to the
local-moment system is a well-known feature of the
\textit{``concentrated''} ferromagnetic semiconductors. Striking
consequences of this special temperature dependence are the \textit
{``red shift''} of the optical absorption
edge \cite{Wachter79} and the 
metal-insulator transition in Eu-rich EuO \cite{Penney72,SinNo03}. The
responsible exchange interaction appears to be decisive for the physics
of the DMS, too. It creates the ferromagnetism in these materials. An
important question is whether and how the disorder of the localized
 magnetic (Mn$^{2+}$)
moments influences the magnetic stability. With respect to the main
goal, namely, reaching room temperature ferromagnetism, the disorder
aspect has to be considered as a central point to clarify.

The natural precondition for an understanding of the
\textit{``diluted''} ferromagnetic semiconductors is to have understood
the \textit{``concentrated''} counterparts. From a theoretical point of
view, that means to find a convincing (approximate) solution of the
(ferromagnetic) KLM
\cite{nrrmk03,SaNo02,schiller01:_kondo,NRMJ97}. The general solution of
the sophisticated many-body problem provoked by KLM is not yet
available. The model describes the mutual influence of two well-defined
electronic subsystems, localized magnetic moments and itinerant band
electrons. It turns out to be a non-trivial challenge to  treat both
subsystems simultaneously on the same theoretical level. To our 
information, such a theory does not
yet exist. It is the aim of this paper to propose a new way to approach
this  problem.

The second step is to introduce disorder of the localized magnetic
moments by dilution and to inspect its influence on the magnetic
stability \cite{Timm03,TSO02,BKB03}.
 Does
the disorder weaken or even strengthen the ferromagnetism? How can we
understand the fact that surprisingly low moment concentrations and carrier
densities are able to mediate a ferromagnetic ordering in diluted
magnetic semiconductors Ga$_{1-x}$Mn$_{x}$As. The final goal is to work
out the prerequisites for room temperature ferromagnetism in diluted
magnetic semiconductors. We therefore, derive the magnetic phase diagram
of a diluted Kondo-lattice (concentration $x$) in terms of model
parameters such as $x$, the carrier concentration $n\le x$, and the
exchange coupling $J$. For this purpose, we introduce in the next section
the KLM and a proposal for its electronic
selfenergy. The concept of disorder is developed in section 3, while
the magnetic phase diagram (Curie temperature $T_{C}=T_{C}(x,n,J)$) is
read off from the singularities of the paramagnetic susceptibility
(section 4). The results are discussed in section 5.

\section{Kondo-lattice model}
The (ferromagnetic) Kondo-lattice model is today certainly one of the
most frequently applied models in solid state theory, because of its
great variety of potential applications to technologically promising
topics in the wide field of collective magnetism. It refers to magnetic
materials that get their magnetic properties from a system of localized
magnetic moments being indirectly coupled via interband exchange to
itinerant conduction electrons. Many characteristic features of such
materials can be traced back to this interband exchange.The respective
 model-Hamiltonian \cite{nrrmk03,SaNo02} 
\begin{equation}
  \label{eq:KLM}
  H=H_{s}+H_{sf}
\end{equation}
describes the interaction of itinerant band electrons in a homogeneous
magnetic field $B$ ($\mu_{B}$: Bohr magneton),
\begin{equation}
  \label{eq:sel}
  H_{s}=\sum_{ij\sigma}\left(T_{ij}-z_{\sigma}\mu_{B}B\delta_{ij}\right)c_{i\sigma}^{\dagger}c_{j\sigma}
\end{equation}
and localized magnetic moments (spins $\mathbf{S}_{i}$) via an
intraatomic exchange:
\begin{equation}
  \label{eq:sfww}
  H_{sf}=-J\sum_{j}\mathbf{\sigma}_{j}\cdot \mathbf{S}_{j}=-\frac{1}{2}J\sum_{j\sigma}(z_{\sigma}S_{j}^{z}n_{j\sigma}+S_{j}^{-\sigma}c_{j\sigma}^{\dagger}c_{j-\sigma})
\end{equation}
without any direct exchange interaction between the localized spins.
 $c^{\dagger}_{j\sigma}$
($c_{j\sigma}$) is the creation (annihilation) operator for a Wannier
electron with spin $\sigma$ ($\sigma=\uparrow$, $\downarrow$) at site
$\mathbf{R}_{j}$ ($n_{j\sigma}=c_{j\sigma}^{\dagger}c_{j\sigma}$;
$z_{\sigma}=\delta_{\sigma\uparrow}-\delta_{\sigma\downarrow}$;
$S_{j}^{\sigma}=S_{j}^{x}+{\rm i}z_{\sigma}S_{j}^{y}$). $J$ is
the exchange coupling and $T_{ij}$ the hopping integral. The latter is
connected by Fourier transformation to the Bloch energy
$\epsilon(\mathbf{k})$:
\begin{equation}
  \label{eq:hop}
  T_{ij}=\frac{1}{N}\sum_{\mathbf{k}}\epsilon(\mathbf{k}){\rm e}^{{\rm i}
    \mathbf{k}\cdot(\mathbf{R}_{i}-\mathbf{R}_{j})}
\end{equation}
In spite of its simple structure, the model-Hamiltonian (\ref{eq:KLM})
provokes a rather sophisticated many-body problem, which, at least for
the general case, could not be solved exactly
up to now. One of the main challenging questions is whether or not and
under what conditions the interband exchange $J$ may cause a collective
(ferromagnetic) ordering of the coupled local-moment/ itinerant electron
system. Conventional second-order perturbation theory predicts an
indirect Heisenberg exchange (Rudermann-Kittel-Kasuya-Yoshida
(RKKY)) between the local moments. Approximate Statistical Mechanics of
the resulting Heisenberg model, e.g. in the frame work of the Tyablikov
method \cite{BT59}, indeed predicts ferromagnetism, but only for very low band
occupations $n=\frac{1}{N}\sum_{j\sigma}\langle
n_{j\sigma}\rangle$ ( Ref. \onlinecite{SaNo02}). A \textit{modified} RKKY theory
presented in Ref. \onlinecite{SaNo02}, which takes into account higher order terms of the
induced conduction electron spin polarization by a mapping of the s-f
interaction (\ref{eq:sfww}) on an effective Heisenberg-Hamiltonian,
results in a magnetic phase diagram with respect to the coupling
strength $J$ and the band occupation $n$. To our information, however, there
does not exist a complete theory that treats the electronic part 
 and the magnetic moment part of the KLM on the same level and 
in the same theoretical framework. 
Admittedly, this indeed appears to be a
rather involved task. Very often, only the electronic problem is
investigated while the local moment magnetization is phenomenologically 
simulated by a Brillouin function \cite{nrrmk03,nrrm01,HiNo04}
. Such procedure presumes ferromagnetism, that by no means is always valid,
 without deriving it
selfconsistently within the KLM. 

The electronic part of the many-body problem is solved as soon as the
single-electron Green function $G_{\mathbf{k}\sigma}(E)$ is available or,
equivalently, the electronic selfenergy $M_{\sigma}(E)$:
\begin{equation}
  \label{eq:Gf}
  G_{\mathbf{k}\sigma}(E)=\frac{\hbar}{E-\epsilon(\mathbf{k})+z_{\sigma}\mu_{B}B-M_{\sigma}(E)}
\end{equation}
For simplicity, we assume from the very beginning a wave-vector independent
 selfenergy. A
$\mathbf{k}$-dependence of the selfenergy would be mainly due to magnon
energies $\hbar\omega(\mathbf{k})$ appearing as a consequence of magnon
emission and absorption processes by the band electron \cite{NRMJ97}. However, the
neglect of a direct Heisenberg exchange between the localized spins in
the KLM (\ref{eq:KLM}) can be interpreted as the
$\hbar\omega(\mathbf{k})\rightarrow 0$-limit. In a previous 
paper\cite{nrrm01}, we have developed a theory for the electronic selfenergy,
which fulfills, in the low carrier-density limit ($n\rightarrow0$),
 all the known exact limiting cases:
\begin{eqnarray}
  \label{eq:SE}
  M_{\sigma}(E)&=&-\frac{1}{2}Jz_{\sigma}\langle S^{z}\rangle \nonumber\\
   &+&\frac{1}{4}J^{2}\frac{a_{\sigma}G_{0}(E-\frac{1}{2}Jz_{\sigma}\langle
    S^{z}\rangle
-z_{\sigma}\mu_{B}B)}{1-b_{\sigma}G_{0}(E-\frac{1}{2}Jz_{\sigma}\langle
S^{z}\rangle-z_{\sigma}\mu_{B}B)} \qquad
\end{eqnarray}
An extensive discussion of the
reliability of this selfenergy can be found in the above mentioned paper\cite{nrrm01}.
 $a_{\sigma}$, $b_{\sigma}$ are parameters which are fixed by rigorous
high-energy expansions to fulfill the first four spectral moments:
\begin{equation}
  \label{eq:par}
  a_{\sigma}=S(S+1)-z_{\sigma}\langle S^{z}\rangle(z_{\sigma}\langle
  S^{z}\rangle+1);\ \  b_{\sigma}=b_{-\sigma}=\frac{J}{2}
\end{equation}
$G_{0}(E)$ is the \textit{``free''} propagator:
\begin{equation}
  \label{eq:prop}
  G_{0}(E)=\frac{1}{N}\sum_{\mathbf{k}}\frac{1}{E-\epsilon(\mathbf{k})
}
\end{equation}
Since Eq.(\ref{eq:SE}) is exact for a maximum number of special cases in
the low-density limit, it should represent a reasonable starting point
for the description of ferromagnetic semiconductors, which, by definition,
is restricted to low densities of itinerant charge
carriers. $M_{\sigma}(E)$ is the electronic selfenergy for the
\textit{``concentrated''} (periodic) Kondo lattice. In the next section,
we propose how to model the disorder of the magnetic moments in
diluted ferromagnetic semiconductors.

\section{Electronic selfenergy of the diluted system}
We consider a binary alloy of constituents $\alpha$ (concentration $1-x$)
and $\beta$ (concentration $x$). $\alpha$ symbolizes nonmagnetic sites 
(Ga$^{3+}$), while site $\beta$
carries a magnetic moment (Mn$^{2+}$ ion) being exchange coupled via
(\ref{eq:sfww}) to the itinerant charge carriers. The atomic level of $\alpha$ sites is
in the presence of a magnetic field $B$:
\begin{equation}
  \label{eq:Aatom}
  \epsilon_{\alpha\sigma}=T_{0}-z_{\sigma}\mu_{B}B
\end{equation}
On $\beta$ sites, however, the local interband exchange $H_{sf}$
(\ref{eq:sfww}) acts on the charge carriers. That is accounted for by a
\textit{``dynamic''} atomic energy level incorporating the selfenergy
$M_{\sigma}(E)$ (\ref{eq:SE}):
\begin{equation}
  \label{eq:Batom}
 \epsilon_{\beta\sigma}=T_{0}+M_{\sigma}(E)-z_{\sigma}\mu_{B}B 
\end{equation}
We consider the charge carriers in the \textit{``diluted''} Kondo lattice as a system
of particles propagating in the above-defined fictitious binary
$\alpha\beta$-alloy, thereby neglecting a Coulomb disorder potential
which might be important in some circumstances\cite{Timm03}
(e.g. metal-insulator transition). 
The single-particle properties can then be derived from the
propagator
\begin{equation}
  \label{eq:Rprop}
  R_{\sigma}(E)=\int\limits_{-\infty}^{+\infty}
    \!\!{\rm d}\omega\, \frac{\rho_{0}(\omega)}{E-\omega-\Sigma_{\sigma}(E)}
\end{equation}
where $\Sigma_{\sigma}(E)$ is now the electronic selfenergy in the
diluted system and
$\rho_{0}(x)$ the Bloch-density of states of the non-interacting
carriers. For the determination of the decisive selfenergy we use a
standard CPA formalism \cite{Cpa-Elliott}, i.e., this quantity
is determined by the CPA equation:
\begin{eqnarray}
 \label{eq:CPA}  
  0 &=&(1-x)\frac{-z_{\sigma}\mu_{B}B-\Sigma_{\sigma}(E)}{1-R_{\sigma}(E)
  (-z_{\sigma}\mu_{B}B-\Sigma_{\sigma}(E))}\nonumber\\
  &+&x\frac{M_{\sigma}(E)-z_{\sigma}\mu_{B}B-\Sigma_{\sigma}(E)}{1-R_{\sigma}(E)
  (M_{\sigma}(E)-z_{\sigma}\mu_{B}B-\Sigma_{\sigma}(E))} \qquad
\end{eqnarray}
The limiting cases $x=0$ ($\Sigma_{\sigma}(E)=-z_{\sigma}\mu_{B}B$) and
$x=1$ (\textit{``concentrated''} KLM with
$\Sigma_{\sigma}(E)=M_{\sigma}(E)-z_{\sigma}\mu_{B}B$) are obviously
fulfilled.

The  configurational averaging, inherent in CPA, takes care for
translational symmetry and therewith for site-independent average
spin-dependent occupation numbers:
\begin{equation}
  \label{eq:TZ}
  \langle n_{\sigma}\rangle
   = \int\limits_{-\infty}^{+\infty} \!{\rm d}E\,
     \frac{ \rho_{\sigma}(E) }{{\rm e}^{\beta(E-\mu)} + 1} 
   \,\equiv
     \int\limits_{-\infty}^{+\infty} \!{\rm d}E\,
  f_{-}(E)\rho_{\sigma}(E)
\end{equation}
\( f_{-}(E) \) is the Fermi function, 
$\mu$ the chemical potential,  and $\rho_{\sigma}(E)$ the
quasiparticle density of states of the {\it interacting} particle system:
\begin{equation}
  \label{eq:QDOS}
  \rho_{\sigma}(E)=-\frac{1}{\pi}{\rm Im}R_{\sigma}(E)
\end{equation}

In the special case of a paramagnetic system and $B\rightarrow 0^{+}$
(``pm'') the density of states reads
\begin{equation}
  \label{eq:PQDOS}
 \rho_{\rm pm}(E)=-\frac{1}{\pi}{\rm Im}R_{\rm pm}(E) 
\end{equation}
with
\begin{equation}
  \label{eq:PRprop}
 R_{\rm pm}(E)=\int\limits_{-\infty}^{+\infty}
 \!{\rm d}\omega\, \frac{\rho_{0}(\omega)}{E-\omega-\Sigma_{\rm pm}(E)}  
\end{equation}
The paramagnetic selfenergy obeys the CPA equation (\ref{eq:CPA}) in the
following form:
\begin{eqnarray}
  \label{eq:PCPA} 
  0&=&(1-x)\frac{-\Sigma_{\rm pm}(E)}{1+R_{\rm pm}(E)\Sigma_{\rm pm}(E)} \nonumber\\
 &+&x\frac{M_{\rm pm}(E)-
 \Sigma_{\rm pm}(E)}{1-R_{\rm pm}(E)(M_{\rm pm}(E)-\Sigma_{\rm pm}(E))}\qquad
\end{eqnarray}
The selfenergy for the paramagnetic phase of the
\textit{``concentrated''} KLM (\ref{eq:SE}) becomes 
in view of Eq. (\ref{eq:par})
especially simple:
\begin{equation}
  \label{eq:parSE}
  M_{\rm pm}(E)=\frac{1}{4}J^{2}\frac{S(S+1)G_{0}(E)}{1-\frac{1}{2}JG_{0}(E)}
\end{equation}
 We need these expressions when calculating, in the next section, the
paramagnetic susceptibility of the itinerant charge carriers. 

\section{Static magnetic susceptibility}
In the theory of Ref. \onlinecite{nrrm01}, the local moment magnetization
$\langle S^{z}\rangle$ is left as a parameter which was represented by
a Brillouin function. However, for a given parameter constellation, it is
 by no means predetermined that the
system will indeed be ferromagnetic,
i. e. a full theory would require a self-consistent treatment of
$\langle S^{z}\rangle$ within the (\textit{``concentrated''} or
\textit{``diluted''}) KLM. This turns out to be a rather
non-trivial goal. For our purpose, to derive the magnetic phase
diagram of the KLM, we circumvent this problem by exploiting the static
susceptibility of the itinerant electron subsystem:
\begin{equation}
  \label{eq:susc}
  \chi(T)=\sum_{\sigma}z_{\sigma}\left(\frac{\partial}{\partial B}\langle
    n_{\sigma}\rangle\right)^{B\rightarrow 0}_{T>T_{C}}
\end{equation}
We inspect exclusively the possibility of ferromagnetism, the average
occupation number $\langle n_{i\sigma}\rangle$ is therefore
site-independent (Eq. (\ref{eq:TZ})).

The spontaneous magnetization $\langle S^{z} \rangle$ of the local moment
system and the conduction electron spin polarization $\langle
n_{\uparrow}-n_{\downarrow} \rangle$ are mutually conditional. Therefore,
they become critical for the same parameters, in particular, at the same
temperature. In the critical region, we can therefore assume:
\begin{equation}   \label{eq:ansatz}
   \left(\frac{\partial}{\partial B}\langle S^{z} \rangle\right)_{T>T_{C}}^{B \rightarrow 0}=\eta\cdot\chi(T) 
 \end{equation}
The proportionality of the response functions can be traced back to a 
proportionality of the expectation values $\langle S^{z} \rangle$ and
$\langle n_{\uparrow}-n_{\downarrow} \rangle$, which is in terms of
a Taylor expansion certainly fulfilled. In order to concentrate on the
effects of dilution, we made a simple ansatz for the proportionality 
factor $\eta$, which neglects the dependence on model parameters and
temperature. Instead we assume a equivalence of the reduced quantities
\begin{equation}
  \label{eq:AN}
  \frac{\langle S^{z}\rangle}{S}\Leftrightarrow \frac{\langle n_{\uparrow}-n_{\downarrow}\rangle}{n}
\end{equation}
and take $\eta=\frac{S}{n}$. This ansatz, plausible as it is, can
probably been replaced by more profound theories in an improved
approach. 

A straightforward derivation of the itinerant-electron susceptibility
$\chi$ according to Eqs. (\ref{eq:Rprop}), (\ref{eq:SE}), (\ref{eq:susc}),
and (\ref{eq:ansatz}) eventually ends up with the following expression:
\begin{equation}
  \label{eq:result}
  \chi(T)=-2\mu_{B}\frac{Q_{x}(T)+K_{x}(T)}{1+\eta JK_{x}(T)}
\end{equation}
For clarity, the lengthy derivation of $Q_{x}(T)$ and $K_{x}(T)$ is 
shifted to the appendix.

From the singularities of the paramagnetic susceptibility $\chi$, we find
the Curie temperature $T_{C}$ as function of model parameters such as
lattice structure, spin value $S$, moment concentration $x$, band occupation
$n\leq x$, and exchange
coupling $J$. The singularities are the solutions of the following equation:
\begin{equation}
  \label{eq:BG}
  0=1+\eta JK_{x}(T=T_{C})
\end{equation}
 The instabilities of the paramagnetic
phase towards ferromagnetism are thus given by the solutions of this
equation.

\section{Magnetic phase diagram}
We have evaluated the criterion for ferromagnetism (\ref{eq:BG}) for an
sc lattice where the width $W$ of the Bloch band has been chosen to be 1
eV. The goal is to find out for which parameter constellations (moment
concentration $x$, bandoccupation $n\leq x$, exchange coupling $J$) the
system becomes ferromagnetic and what are the values for the Curie
temperature $T_{C}=T_{C}(x,n,J)$. We start the analysis of the results
with a discussion of the \textit{``concentrated''} systems, where
(having substances like EuO and Gd in mind) the exchange coupling
constant $J$ is ferromagnetic. To be consistent, we have
restricted ourselves even in the case of \textit{``diluted''} systems 
to a ferromagnetic exchange coupling $J>0$, although the most topical 
diluted magnetic semiconductors seem to have an antiferromagnetic 
coupling. Furthermore, our model study considers the coupling of
electrons to localized moments, the case of holes instead of electrons
will not essentially change the important statements. 

\begin{figure}[htbp]
  \centering
  \epsfig{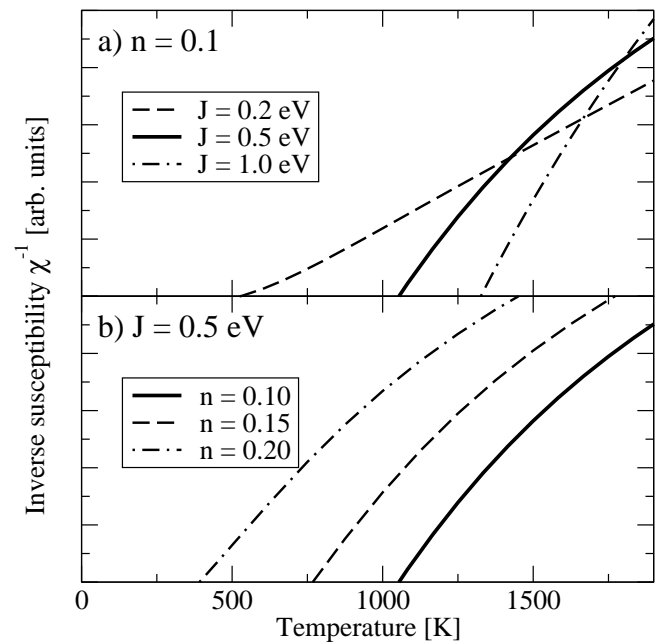}
  
  \caption{Paramagnetic inverse susceptibility of the
    \textit{``concentrated''} $(x=1)$ Kondo lattice as function of the
    temperature, in (a) for a fixed band occupation $n=0.1$ and
    different exchange couplings, in (b) for a fixed exchange coupling 
    $J=0.5 {\rm eV}$ and different carrier concentrations $n$.}
  \label{fig:CurieWeiß}
\end{figure}
Let us first inspect the case of the \textit{``concentrated''} Kondo
lattice ($x=1$). Fig. \ref{fig:CurieWeiß} shows the paramagnetic
inverse susceptibility of the band electrons as function of the temperature
for various parameter constellations ($n$,$J$). For sufficiently high
temperatures and almost all parameter constellations, a Curie-Wei{\ss}
behaviour can be recognized. From the zeros of $\chi^{-1}$ we can read
off the respective Curie temperature. In some cases two zeros are found
(not shown in the figure). 
The requirement that $\chi$ must be positive in the paramagnetic phase
($T>T_{C}$) makes the choice of the physically relevant solution unique. 

\begin{figure}[thbp]
  \begin{center}
        \epsfig{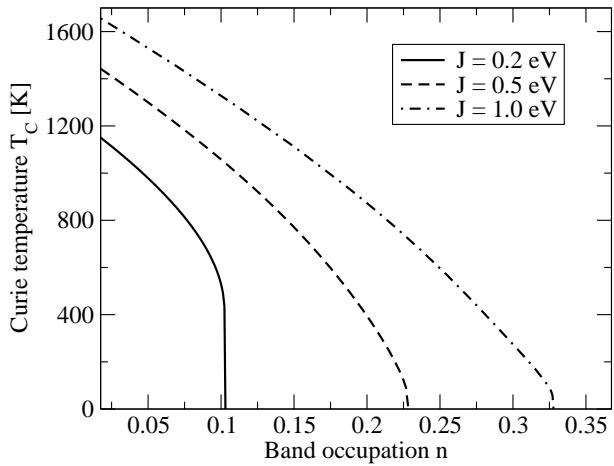}
  
        \caption{Curie temperature as a function of the band occupation $n$
      for various exchange couplings $J$ in the \textit{``concentrated''}
      ($x=1$) Kondo-lattice
      model. Parameters: sc lattice, $W=1 {\rm eV}$, $S=\frac{5}{2}$}
 \label{fig:phasediagram}
 \end{center}
\end{figure}

The band occupation $n$ enters the susceptibility (\ref{eq:susc}) and 
therefore the calculated $T_C$ via the chemical potential $\mu$, which
is accordingly determined with the help of Eq. (\ref{eq:TZ}). 
Additionally $n$ is included in the choice of $\eta$. 
Fig. \ref{fig:phasediagram} demonstrates that ferromagnetism does exist
with a distinct band occupation dependence of the Curie temperature. The most
remarkable feature is the restriction of ferromagnetism to surprisingly
low carrier concentrations $n$. Arbitrarily small band occupations are
sufficient to create a ferromagnetic order. In any case, the Curie temperature is
zero for $n=0$. It was, however, numerically not possible to decide
whether or not there is a steep but continuous increase to finite
values. Note that the KLM does not consider a
direct exchange between the localized moments. So the collective
ordering is fully mediated by the interband exchange, i. e. by the conduction
 electron spin polarization. 
The width of the ferromagnetic phase
on the $n$ axis increases with the exchange coupling strength $J$, being
restricted, however, even for strong couplings to low itinerant electron
concentrations. The maximum value of the Curie temperature also increases with
$J$. Typical $J$ values for (``concentrated'') ferromagnetic semiconductors such
 as EuO and EuS
are of the order of some tenth of eV (Ref. \onlinecite{schiller01:_temper_euo,MuNo02}).

Similar results are found with the
``modified'' RKKY of Refs. \onlinecite{SaNo02,NRMJ97}, where an effective
Heisenberg model is solved by the Tyablikov approximation
\cite{BT59}. The model theory in Ref. \onlinecite{ChaMil01} yields also qualitatively
the same $T_{C}$ behaviour, namely a steep increase of $T_{C}$ for
very weak band occupations, a rather distinct maximum and then also a
very rapid decrease to zero. The new feature of our theory (Fig.
\ref{fig:phasediagram}) is the $T_{C}$ behaviour for $n \rightarrow 0$.

\begin{figure}[thbp]
        \epsfig{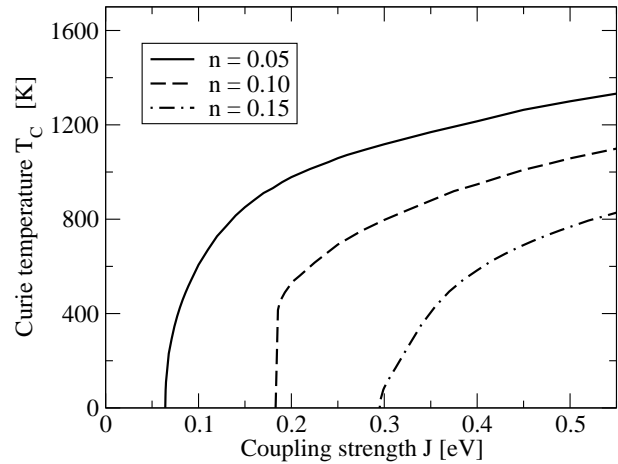}
        \begin{center}
                    \caption{Curie temperature as a function of the
                      exchange coupling strength J for three different
                      band occupations n in the
                      \textit{``concentrated''} ($x=1$) Kondo-lattice model.
                       Parameters: sc lattice, $S=\frac{5}{2}$, $W=1 {\rm eV}$}
          \label{fig:Jdep}
        \end{center}
      \end{figure}

The general $J$ dependence of $T_{C}$ is shown in Fig. \ref{fig:Jdep}.
Two features are worth mentioning. Firstly, $T_{C}$ appears to run
into a saturation in the strong coupling region. This is similar to what
is reported in Ref. \onlinecite{SaNo02}. In the present theory, however, the 
saturation needs a substantially stronger exchange
coupling. Secondly, a critical $J=J_{c}(n)$ is needed to switch on
ferromagnetism, which, at least in the low concentration regime,
increases with increasing $n$.

\begin{figure}[htbp]
  \centering
  \epsfig{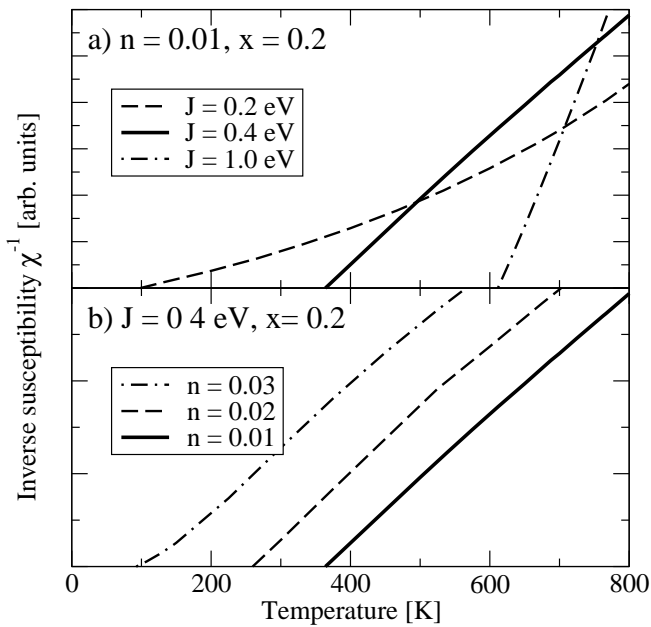}
 
  \caption{Paramagnetic inverse susceptibility of the
    \textit{``diluted''} Kondo lattice as function of the
    temperature, in (a) for a fixed band occupation $n=0.01$ and
    different exchange couplings, in (b) for a fixed exchange coupling $J=0.4 {\rm eV}$ and different carrier concentrations $n$.}
  \label{fig:CurieWeißDil}
\end{figure}
We now inspect the influence of the dilution of the moments ($x<1$). We
assume that each magnetic ion can in principle donate one electron to the
conduction band. However, not all these charge carriers can be
considered as really
itinerant, so that $n\le x$. Therewith we simulate the situation in the
diluted ferromagnetic semiconductors. In the case of Mn$^{2+}$ in 
Ga$^{3+}$As$^{3-}$, e.g., holes are created in the GaAs valence band 
which are partly
compensated by antisites \cite{Dietl02}. The inspection of the
paramagnetic susceptibility as function of temperature for a given
parameter constellation (Fig. \ref{fig:CurieWeißDil}) makes it clear that
ferromagnetism does exist in the diluted moment system, too. The
\begin{figure}[htbp]
  \begin{center}
    \epsfig{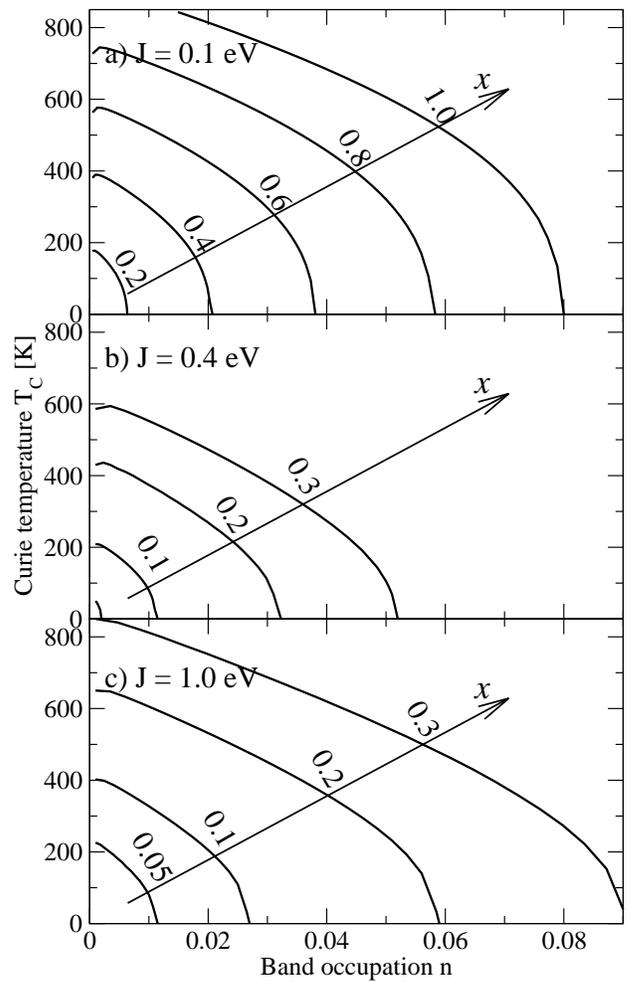}
    \caption{Curie temperature as a function of the band occupation $n$
      for various concentrations $x$ of magnetic moments in the 
      \textit{``diluted''}($x<1$) Kondo-lattice
      model. (a) $J=0.1 {\rm eV}$, (b) $J=0.4 {\rm eV}$, 
      (c) $J=1.0 {\rm eV}$. Parameters: sc lattice, 
      $W=1 {\rm eV}$, $S=\frac{5}{2}$}
 \label{fig:phasediagramDil}
 \end{center}
\end{figure}
resulting Curie temperature is plotted in Fig.
\ref{fig:phasediagramDil} as function of the carrier concentration $n$
for various moment concentrations $x$. As in the case of the
\textit{``concentrated''} system, ferromagnetism is restricted to the very
low concentration region. Also the $J$ dependence of the Curie 
temperature for a given $(x,n)$ pair is very similar to that for 
the \textit{``concentrated''} systems plotted in Fig. \ref{fig:Jdep}.  
What is remarkable, however, is the fact that the concentration
$n$ must be very much smaller than the concentration $x$ in order to
allow ferromagnetic ordering. The compensation effects observed in
diluted magnetic semiconductors (antisites,...) seem to be a necessary
precondition for ferromagnetism in the diluted system. For $n=x$
ferromagnetism is excluded. An explanation for this is given by 
the quasiparticle density of states.

\begin{figure}[htbp]
  \centering
  \epsfig{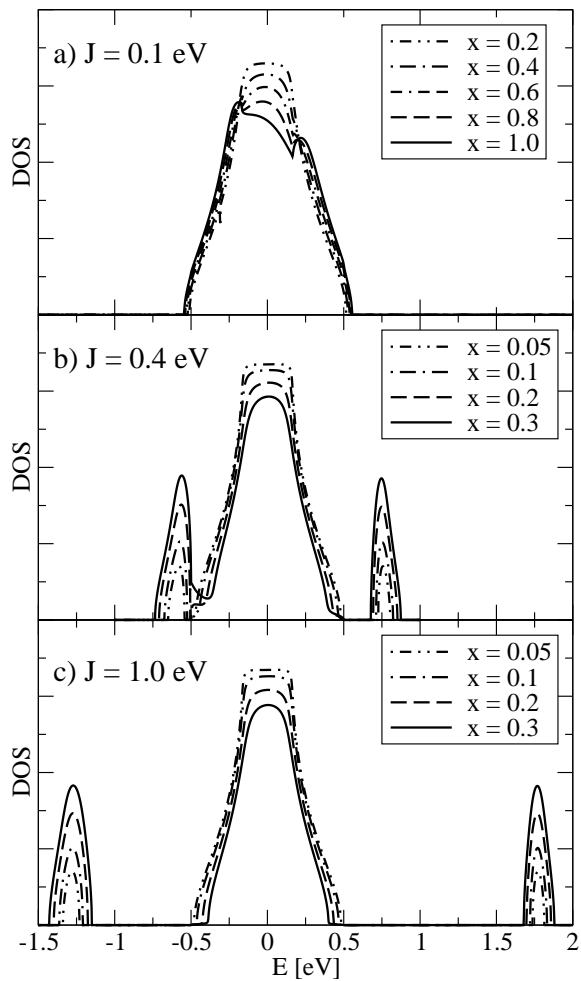}
  \caption{Paramagnetic quasiparticle density of states of the 
  \textit{``diluted''} Kondo-lattice
    model in the paramagnetic phase as function of energy for different
    values of the moment concentration $x$ and three different exchange
    couplings $J$. Parameters: sc lattice, $W=1 {\rm eV}$, $S=\frac{5}{2}$}
  \label{fig:QDOS}
\end{figure}

For sufficiently high values of $J$, the (paramagnetic) quasiparticle 
density of
states (Fig. \ref{fig:QDOS}) consists of three parts. The low-energy
and the high-energy subbands are built up by states from the correlated
$\beta$-sites, while the middle structure is due to the uncorrelated
$\alpha$-sites. The correlated subbands, which are exclusively responsible
 for a possible
magnetic order, are exchange-split by about $\frac{1}{2}J(2S+1)$. When
the three structures are well separated, then, the area under the two
correlated peaks amounts to $x$ while that of the uncorrelated middle band
is $1-x$. With increasing $x$, i. e. higher moment density, more and more
spectral weight is shifted into the correlated quasiparticle
subbands. In simple terms, the two correlated bands can be understood as
follows: An electron propagating in the low-energy subband hops mainly
over lattice sites where it can orient its spin parallel to the
local-moment (Mn$^{2+}$) spin ($\sim -\frac{1}{2}JS$). In the
high-energy subband, the spin orientation is predominantly antiparallel
($\sim +\frac{1}{2}J(S+1)$). Since we have used for the selfenergy
$M_{\sigma}(E)$ the low-density approach of Ref. \onlinecite{nrrm01}, the QDOS
does not exhibit a noteworthy band occupation dependence.

First precondition for ferromagnetism is that the Fermi edge lies in one of
the correlated subbands. We observe in principle the same general structure as 
in the concentrated case ($x=1$) exhibited
in Fig. \ref{fig:phasediagram}. Extremely low carrier concentrations are
already sufficient to induce ferromagnetism. Roughly estimated, we find
ferromagnetic ordering for bandoccupations $0 < n< n_{c}(J)\cdot x$,
where $n_c(J)$ is the critical bandoccupation for $x=1$ at a given $J$. 

It is indeed observed for diluted magnetic semiconductors that the 
number of itinerant carriers is substantially smaller than the number 
of local moments\cite{Dietl02}. In Ga$_{1-x}$Mn$_{x}$As, e.g., each 
Mn$^{2+}$ ion in principle provides one hole in the valence band. 
However, only a certain percentage of them are really itinerant, the 
others are compensated, e.g., by antisites or interstitial Mn atoms, 
that act as donors. Erwin and Petukhov \cite{EP02} were the first
to suggest that such compensation effects might be in favour
of a collective order. In the limit $J \to \infty$ they mapped the 
Hamiltonian (\ref{eq:KLM}) on an effective Heisenberg model and 
evaluated the latter using classical percolation theory. 
With our treatment of the Kondo-lattice model, which is valid for
quantum spins and finite $J$, we can confirm that compensation is
necessary for the existence of ferromagnetism. The reason is the
complete filling of the lower correlated subband in Fig. \ref{fig:QDOS} 
for $n=x$. This corresponds in the \textit{``concentrated''} local-moment
systems (Fig. \ref{fig:phasediagram}) to a half-filling of the 
correlated spectrum, which is known to prevent a magnetic order
\cite{SaNo02,NRMJ97}. In contrast to Erwin et al. the $n_{c}(J)$
determined from our results is substantially smaller than $x$. More 
recently a similar behaviour was found by Bouzerar et al.\cite{BKB03} 
and Brey et al.\cite{BGS03}. 

Our findings are in particular interesting, because they seem to be 
in disagreement with some ab initio calculations\cite{SH01,KTD04}. 
These papers mostly refer to compensation effects of As 
antisites. Since interstitial Mn atoms have a different magnetic 
behaviour, its compensation might have a different effect 
on $T_C$, too. Nevertheless, this point is apparently 
still an exciting open question, both for experimentalists and 
theoreticians.

\section{Summary}
In conclusion, it can be stated that the basic theory for the selfenergy
(\ref{eq:SE}) is undoubtedly justifiable for the low-concentration limit
of the KLM. Fortunately, this is obviously just the most relevant region
for stable ferromagnetism \cite{SaNo02,NRMJ97,ChaMil01}. The assumption
of equivalent criticality (\ref{eq:ansatz}) of the two subsystems of the
KLM is certainly acceptable, while the choice of the parameter $\eta$
(see Eq.(\ref{eq:AN})) seems to be plausible. Nevertheless, the latter surely
needs stronger confirmation. Interesting remarks about this fact can 
be found in Ref. \onlinecite{TOH04}. A change of $\eta$, however, does not
qualitatively alter the findings of the theory. The absolute values of the
Curie temperatures depend of course sensitively on $\eta$.

We have shown by a CPA-type treatment of the disordered KLM how the
magnetic disorder in diluted local-moment systems influences the
existence of a ferromagnetic phase and  the respective Curie
temperature. The model study gives a qualitative explanation of the
ferromagnetism in diluted magnetic semiconductors. A main consequence of
our model study is that a substantial
compensation of the itinerant charge carriers ($n<x$) by antisites or
other mechanisms appears to be a necessary condition for the existence
of a ferromagnetic ordering. It is intended for the future to apply our
theory to real diluted magnetic semiconductors (negative $J$!).
   
\begin{acknowledgments}
Financial support by the \textit{``Volkswagenstiftung''} is gratefully
acknowledged. This work also benefitted from the
support of the Sonderforschungsbereich 290 of the Deutsche 
Forschungsgemeinschaft.
\end{acknowledgments}

\appendix

\section{}

We give here the full analytical solution for the paramagnetic
susceptibility (\ref{eq:result}). By definition (\ref{eq:susc})
it is determined by the electron polarisation.
Substituting (\ref{eq:Rprop}) and (\ref{eq:QDOS}) into the
spectral theorem (\ref{eq:TZ}) yields:
\begin{eqnarray}
  \label{eq:AppA1}
  \frac{\partial \langle n_{\sigma} \rangle}{\partial B}
  &=& \int\limits_{-\infty}^{+\infty} \! {\rm d}E \,f_-(E) \times \\
  &&\times\left(-\frac{1}{\pi}{\rm Im}
    \int\limits_{-\infty}^{+\infty} \! {\rm d} \omega \,
    \frac{ \rho_0(\omega) \cdot 
      \left( \frac{ \partial}{\partial B} \Sigma_\sigma(E) \right)}
      {[E-\omega-\Sigma_\sigma(E)]^2}    \nonumber
  \right)
\end{eqnarray}
According to the chain rule the derivative is reduced to that of
\( \frac{ \partial}{\partial B} \Sigma_\sigma(E) \). It is derived
from the application of \( \frac{ \partial}{\partial B} \) to
Eq. (\ref{eq:CPA}). Afterwards the limit $B \to 0$ ist taken. 
Those terms which are proportional to 
\( \frac{ \partial \langle S^z \rangle }{\partial B} = \eta 
\cdot \chi(T) \) give rise to \( K_{x}(T) \), the most important
term of Eq. (\ref{eq:result}).
\begin{eqnarray}
  \label{eq:Denom}  
   K_{x}(T)&=&\int\limits_{-\infty}^{+\infty}
   \! {\rm d}E \, f_{-}(E)\times \nonumber\\
  &&\times\left(-\frac{1}{\pi}{\rm Im} \left[D_{\rm pm}(E)
    \frac{B_{x}(E)H(E)}{N_{x}(E)}\right] \right) \qquad
\end{eqnarray}
The remaining terms are summed to
\( Q_{x}(T) + K_{x}(T) \) with
\begin{eqnarray}
  \label{eq:Num}
 && \hspace{-1em}Q_{x}(T)=\int\limits_{-\infty}^{+\infty}
  \! {\rm d}E\, f_{-}(E)\times \\
 && \times\left(-\frac{1}{\pi}{\rm Im}\left[
 \frac{D_{\rm pm}(E)}{N_{x}(E)}\left(A_{x}(E)-\frac{B_{x}(E)}{1-\frac{1}{2}JG_{0}(E)}
 \right)\right]\right) \nonumber
\end{eqnarray}
Eq. (\ref{eq:result}) is a consequence of the result
\begin{equation}
  \label{eq:PreResult}
  \chi(T) = - J \eta \chi(T) K_{x}(T) -2\mu_{B} [Q_{x}(T)+K_{x}(T)]
\end{equation}
  
In these expressions we have used further abbreviations, which
are chosen according to mathematical simplicity. Hence, the 
individual terms do not carry a particular physical meaning.
The $\omega$ integrations in (\ref{eq:AppA1}) and (\ref{eq:SE})
are denoted as:
\begin{eqnarray}
  \label{eq:Dpm}
   D_{\rm pm}(E)&=&\int\limits_{-\infty}^{+\infty}
   \! {\rm d} \omega\, \frac{\rho_{0}(\omega)}
   {(E-\omega-\Sigma_{\rm pm}(E))^{2}} \\
  \label{eq:Ge1}
  G_{k}(E)&=&\int\limits_{-\infty}^{+\infty}
   \! {\rm d}\omega\, \frac{\rho_0(\omega)}{(E-\omega)^{k+1}} 
\end{eqnarray}
From the variety of terms emerging after differentiating  
Eq. (\ref{eq:CPA}) an $x$-independent factor
\begin{eqnarray}
  \label{eq:HvonE}
  H(E)&=&\frac{1-\frac{1}{2}JG_{0}(E)-\frac{1}{4}J^{2}S(S+1)G_{1}(E)}{(1-\frac{1}{2}JG_{0}(E))^{2}} \qquad
\end{eqnarray}
can be separated. The remaining terms are
\begin{eqnarray}
  \label{eq:NxvonE}
 N_{x}(E)&=&(1-x)\frac{1-D_{\rm pm}(E)\Sigma_{\rm pm}^{2}(E)}{(1+R_{\rm pm}(E)
\Sigma_{\rm pm}(E))^{2}}+ \nonumber\\
&+&x\frac{1-D_{\rm pm}(E)(M_{\rm pm}(E)-\Sigma_{\rm pm}(E))^{2}}{(1-R_{\rm pm}(E)(M_{\rm pm}(E)-
\Sigma_{\rm pm}(E)))^{2}} \qquad\\
  \label{eq:AxvonE}  
  A_{x}(E)&=&\frac{1-x}{(1+R_{\rm pm}(E)\Sigma_{\rm pm}(E))^{2}}+ \nonumber\\
  &+&\frac{x}{(1-R_{\rm pm}(E)(M_{\rm pm}(E)-\Sigma_{\rm pm}(E)))^{2}} \\
  \label{eq:BxvonE}
 B_{x}(E)&=&\frac{x}{(1-R_{\rm pm}(E)(M_{\rm pm}(E)-\Sigma_{\rm pm}(E)))^{2}} 
\end{eqnarray}
Obviously, for the concentrated case, where $x=1$ and 
\( M_{\rm pm}(E) = \Sigma_{\rm pm}(E) \), the algebraic
equations have a much simpler structure. 


\bibliographystyle{plain}

\end{document}